\documentclass{aa}  

\usepackage{graphicx}
\usepackage{txfonts}
\usepackage{subcaption}         
\usepackage{lscape}             
\usepackage{placeins}           

\usepackage{multirow}
\usepackage{microtype}
\usepackage[normalem]{ulem}
\usepackage{xcolor}

\usepackage[colorlinks = true,linkcolor = blue,urlcolor  = blue,citecolor = blue,anchorcolor = blue]{hyperref}

\usepackage{orcidlink}

\begin{document}

   \title{Outer-crust equations of state for neutron stars}


%

   \author{P.S. Koliogiannis\orcidlink{0000-0001-9326-7481}\inst{1},\inst{2}\fnmsep\thanks{Corresponding author: pkoliogi@phy.hr}
        \and N. Paar\orcidlink{0000-0002-6673-6622}\inst{1}\fnmsep\thanks{npaar@phy.hr}}

   \institute{Department of Physics, Faculty of Science, University of Zagreb, Bijeni\v cka cesta 32, 10000, Zagreb, Croatia.
   \and Department of Theoretical Physics, Aristotle University of Thessaloniki, 54124 Thessaloniki, Greece.}

   \date{\today}

 
  \abstract
{The equation of state of the outer crust of neutron stars is sensitive to nuclear mass predictions and provides a direct connection to the properties of nuclei throughout the nuclide map, including those beyond experimental reach.}
{We quantify the impact of contemporary nuclear mass models on the composition and thermodynamic properties of the outer crust and assess the consequences for crust-dominated neutron-star configurations near the minimum-mass limit.}
{We constructed four outer-crust equations of state based on the relativistic energy density functional and machine-learning mass model tables. The equilibrium composition of cold catalysed matter in $\beta$-equilibrium was obtained by minimising the Gibbs free energy per nucleon, and the resulting equations of state were implemented in neutron-star structure calculations.}
{The different mass inputs lead to variations in the equilibrium nuclide sequence, distinct last bound nuclei, and moderate shifts in the neutron-drip density. In contrast, the associated thermodynamic properties, as well as the minimum-mass neutron-star configurations, remain closely aligned across the four outer-crust equations of state.} 
{The model dependence of the outer crust is primarily reflected in the detailed nuclide composition and in the precise location of neutron drip. Nevertheless, the considered outer-crust equations of state yield closely consistent predictions for the relevant neutron-star observables, providing a reliable input for stellar modelling.}
   \keywords{dense matter --
             equation of state --
             stars: low-mass --
             stars: neutron --
             stars: rotation}
             
   \maketitle

\nolinenumbers

\section{Introduction}
The outer solid crust of a neutron star, whose structure is governed by the low-density nuclear equation of state (EOS), extends from the stellar surface to the neutron-drip transition and describes the thermodynamic response of cold catalysed matter. In this density regime, matter consists of neutron-rich nuclei arranged in a Coulomb lattice and embedded in a highly degenerate electron gas, with neutrons remaining bound inside the nuclei. The outer-crust EOS is relevant from microscopic and macroscopic perspectives. Microscopically, it determines the equilibrium sequence of nuclei and the onset of neutron drip, both of which depend sensitively on nuclear ground-state masses. Macroscopically, it provides the low-density input required for stellar structure calculations and contributes to the precise determination of low-mass neutron-star configurations, including the minimum stable mass and crust thickness. The outer crust therefore provides a direct probe of nuclear structure far from stability and is a testing ground for nuclear mass models, while a consistent outer-crust EOS remains essential for neutron-star modelling~\citep{Baym-71,Chamel_2008,RevModPhys.89.015007}.

At the densities characteristic of the outer crust, nuclei are expected to arrange themselves in a body-centred-cubic lattice, which minimises the electrostatic lattice energy, while the degenerate electron gas ensures charge neutrality and provides the dominant contribution to the pressure~\citep{Baym-71,Haensel_1989,Haensel_1994,Chamel_2008}. Although the lattice term only represents a small correction to the total energy density, it plays a non-negligible role in determining the equilibrium composition of nuclides. The resulting sequence reflects a competition between nuclear binding, shell effects, electron contribution, and lattice energy: surface effects favour nuclei with large mass numbers, whereas Coulomb effects penalise large proton numbers~\citep{Baym-71,Haensel_1989,Haensel_1990}. As the density increases, the rising electron chemical potential drives successive electron-capture reactions, shifting the equilibrium composition towards increasingly neutron-rich nuclei. This sequence continues towards the neutron-drip line until the neutron-drip density is reached, where neutrons become unbound from nuclei and form a dilute gas surrounding the nuclear clusters. The onset of neutron drip marks the transition from the outer to the inner crust.

A key ingredient in determining the outer-crust composition is the knowledge of nuclear ground-state masses~\citep{Baym-71,Haensel_1989}. At relatively low densities, the equilibrium nuclides are experimentally known, allowing the crustal composition to be determined with little dependence on theoretical mass models. At higher densities, however, the sequence approaches the neutron-drip line, where experimental information on nuclei far from the valley of $\beta$-stability becomes increasingly limited. Theoretical mass models are therefore required to predict the properties of the highly neutron-rich nuclei that may populate the deepest layers of the outer crust. Direct mass measurements remain essential in this context, since they can alter the predicted equilibrium sequence. A notable example is the experimental mass measurement of $^{82}$Zn, which was shown to disfavour its occurrence in the outer crust and to modify the predicted sequence near the $N\simeq50$ shell region~\citep{PhysRevLett.110.041101}.

The EOS of the neutron-star outer crust has been investigated for several decades, following the seminal work of~\citet{Baym-71} [hereafter BPS], who formulated the EOS of cold catalysed matter below the neutron-drip point and determined the corresponding sequence of equilibrium nuclei using the nuclear mass table of~\citet{MYERS19661}. Subsequent studies extended this framework by incorporating updated experimental mass evaluations and a broad range of theoretical nuclear mass models, from empirical mass formulas to more microscopic descriptions. In particular,~\citet{Haensel_1989} employed Hartree--Fock--Bogoliubov calculations with a Skyrme effective interaction to investigate the composition of matter below the neutron drip, while~\citet{Haensel_1994} combined experimental atomic-mass data from~\citet{AUDI19931} with theoretical nuclear mass tables from~\citet{MOLLER1988213} and~\citet{ABOUSSIR1992155}. Relativistic nuclear descriptions have also been explored, ranging from systematic comparisons of relativistic and non-relativistic mass models to dedicated studies of the role of the nuclear symmetry energy and applications of the quark--meson-coupling model~\citep{PhysRevC.73.035804,PhysRevC.78.025807,PhysRevC.102.065801}. Beyond the choice of nuclear masses, several refinements of the BPS treatment have been examined in the literature, including electron exchange and correlation effects, Coulomb screening, finite-size corrections, nuclear deformation, and zero-point motion of the lattice~\citep{PhysRevC.76.065801,Chamel_2008,PhysRevC.83.065810}. More recently, Bayesian and machine-learning techniques have been used to refine nuclear-mass predictions and to quantify the propagation of statistical uncertainties to the outer-crust composition and EOS~\citep{PhysRevC.93.014311,PhysRevC.101.035804,universe7050131}. Using experimental masses from the AME2020 evaluation~\citep{Huang_2021,Wang_2021}, recent calculations have combined the available mass data with different theoretical nuclear mass models and energy-density-functional (EDF) approaches~\citep{PhysRevD.105.043017,Jiang_2025}. In parallel, outer-crust EOSs have been incorporated into unified descriptions of neutron-star matter~\citep{Douchin_2001,Sharma_2015,10.1093/mnras/sty2413}. Bayesian analyses within unified relativistic mean-field frameworks have also recently been employed to constrain the neutron-star crust properties and their associated uncertainties~\citep{rkpk-tny1}.

We construct four outer-crust EOSs and quantify their impact on the neutron-star structure. To this end, we employ three modern nuclear mass tables based on relativistic EDFs: the density-dependent meson-exchange functional DD--ME2~\citep{PhysRevC.71.024312,PhysRevC.89.054320,PhysRevC.91.014324,PhysRevC.93.054310}, and the density-dependent point-coupling functionals DD--PC1~\citep{PhysRevC.78.034318,LIU2024101635} and DD--PCX~\citep{PhysRevC.99.034318,LIU2024101635}. As an independent comparison, we also consider the recently developed machine-learning mass model ELMA~\citep{8n1w-f3b7}, which is based on ensemble learning and model averaging and achieves a root-mean-square error of $\sim$ 65 keV with respect to the AME2020 mass table~\citep{Huang_2021,Wang_2021}.

The paper is organised as follows. In Sect.~\ref{sec:theoretical_framework} we present the theoretical framework used to determine the outer-crust EOSs, while in Sect.~\ref{sec:results} we discuss the resulting properties of neutron stars. Finally, Sect.~\ref{sec:concluding_remarks} summarises the main conclusions.

\section{Theoretical framework}
\label{sec:theoretical_framework}

\subsection{Equilibrium structure of the outer crust}
\label{sec:outer_crust}

To determine the equilibrium composition and EOS of the outer crust, we adopted the formalism of~\citet{Baym-71} for cold catalysed matter, including nuclear-mass and electron corrections discussed by~\citet{PhysRevC.83.065810}. The matter was assumed to form a perfect one-component crystal, with nuclei of mass number $A$ and proton number $Z$ occupying the lattice sites. Charge neutrality requires $n_e=Zn_b/A$, where $n_b$ and $n_e$ denote the baryon and electron number densities, respectively. In the following, we summarise the main formulas.

The total energy per nucleon is
\begin{equation}
    E(A,Z,n_b)
    = \frac{M(A,Z)}{A}
    + \frac{\mathcal{E}_{e}(n_e)}{n_b}
    + \frac{E_{L}(A,Z,n_b)}{A},
\end{equation}
where $M(A,Z)$ is the nuclear mass of the nucleus $(A,Z)$, including the nucleon rest-mass contributions, $\mathcal{E}_{e}(n_e)$ is the electron energy density, and $E_L(A,Z,n_b)$ is the lattice energy per nucleus. Exchange and screening corrections are included in the electron contribution. Additional corrections, such as electron correlations, finite-nuclear-size effects, and the zero-point motion of the nuclei, are neglected because their contributions are negligible in the outer crust~\citep{PhysRevC.83.065810}.

The pressure is dominated by the degenerate electrons, with a smaller contribution from the Coulomb lattice,
\begin{equation}
    P
    = -\left.\frac{\partial (AE)}{\partial V}\right|_{A,Z}
    = \mu_e n_e-\mathcal{E}_{e}(n_e)
    +\frac{n_b}{3}\frac{E_L(A,Z,n_b)}{A},
\end{equation}
where $\mu_{e}$ is the electron chemical potential.

At each pressure, the equilibrium composition and the corresponding EOS were obtained by minimising the Gibbs free energy per nucleon with respect to $A$ and $Z$,
\begin{equation}
    \mu(A,Z,P)\equiv\frac{G}{A}
    =E(A,Z,n_b)+\frac{P}{n_b}.
    \label{eq:gibbs_bps}
\end{equation}
At the transition between adjacent crustal layers, the pressure and the Gibbs free energy per nucleon remain continuous, whereas the baryon density generally changes discontinuously.

In $\beta$-equilibrium, the Gibbs free energy per nucleon is equal to the neutron chemical potential, $\mu_n=G/A$. The neutron-drip point is therefore determined by
\begin{equation}
    \mu_n-m_nc^2=0,
\end{equation}
which marks the onset of neutron emission from nuclei and the appearance of unbound neutrons.

\subsection{Neutron star equations of state}
The nuclear EOS for the description of neutron stars is divided into three density regions: the outer crust, the inner crust, and the core. In the present work, where the main subject is the outer crust and its composition, the formalism and nuclear mass models developed in Sect.~\ref{sec:outer_crust} are employed, providing the equilibrium sequence of nuclei up to the neutron-drip point. For densities $\rho \lesssim 10^{4}~{\rm g~cm^{-3}}$, the EOS of~\citet{PhysRev.75.1561} was used.
For the inner crust, two complementary treatments were considered: (i) a microscopic description based on the SLy interaction~\citep{Douchin_2001}, and (ii) a polytropic parametrisation designed to represent the behaviour of a unified EOS, defined as~\citep{PhysRevLett.83.3362,Carriere_2003} 
\begin{equation}
    P(\mathcal{E}) = A + B \, \mathcal{E}^{4/3},
    \label{eq:poly_fit}
\end{equation}
where $P$ is the pressure, and $\mathcal{E}$ is the energy density. The constants $A$ and $B$ were determined by requiring thermodynamic continuity at the boundaries of the inner crust, namely the neutron-drip point at the outer-crust side and the crust–core transition density obtained from the corresponding liquid core EOS. The choice of Eq.~\eqref{eq:poly_fit} was motivated by the fact that fully microscopic calculations of the inner-crust structure for these specific functionals are not currently available. A consistent microscopic treatment of the inner crust would require large-scale calculations that account for nuclear clusters embedded in a neutron gas, which is beyond the scope of the present work. A fully consistent unified description will be considered in future work.

These approaches allowed us to assess the sensitivity of macroscopic neutron-star observables to the inner-crust modelling and to determine whether the relative differences induced by the outer-crust EOS are preserved when different inner-crust prescriptions are adopted. The uniform liquid core was described using the DD--PC-J31 EOS, based on relativistic EDF with point coupling interaction, consistent with both nuclear and astrophysical constraints~\citep{KOLIOGIANNIS2025139362,Koliogiannis_2026}. The same inner-crust and core EOSs were therefore adopted for all outer-crust models so that the impact of the outer-crust EOS could be examined separately from additional uncertainties associated with the remaining layers.

\subsection{Neutron star structure in the slow-rotation approximation}
The structure of a non-rotating spherically symmetric neutron star is determined by solving the Tolman--Oppenheimer--Volkoff (TOV) equations for a given EOS. The solution provides the radial profiles of the thermodynamic variables and enclosed mass, as well as the gravitational mass and radius of the star~\citep{Shapiro-1983,Glendenning-2000}. To calculate the moment of inertia, we employed the slow-rotation formalism of~\citet{Hartle_1967,Hartle_1968}, retaining terms to first order in the angular velocity. The star was assumed to rotate uniformly with $\Omega\ll\Omega_K$, where $\Omega_K$ is the Keplerian mass-shedding angular velocity. This allowed us to treat rotation as a perturbation of the spherical TOV background.

\section{Results}
\label{sec:results}

\subsection{Crustal equations of state}
\label{sec:eos}
\begin{figure}
    \centering
    \includegraphics[width=\columnwidth]{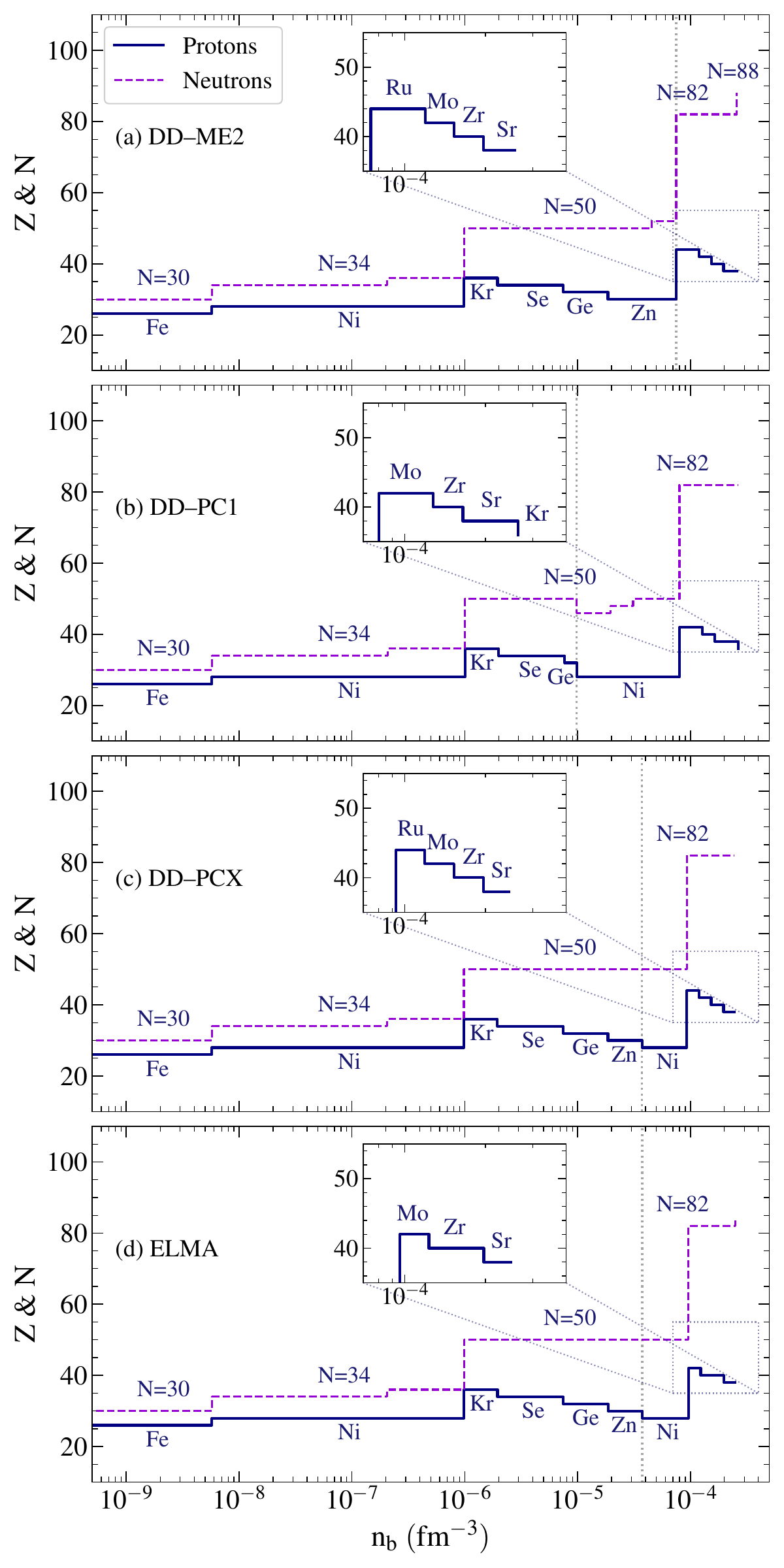}
    \caption{Proton $Z$ and neutron $N$ numbers of the equilibrium nuclei in the outer crust as functions of the baryon density $n_{b}$ for the four nuclear mass models considered: (a) DD--ME2, (b) DD--PC1, (c) DD--PCX, and (d) ELMA. In each panel, the solid (dashed) lines denote the proton (neutron) number. The insets highlight the high-density region close to neutron drip. The vertical dotted lines indicate the extent of experimentally known nuclear masses used in the construction of the EOS~\citep{Huang_2021,Wang_2021}.}
    \label{fig:NZ_vs_nb}
\end{figure}

\begin{table*}
    \centering
        \caption{Neutron-drip properties for the four nuclear mass models considered.}
    \begin{tabular}{lcccccc}
    \hline\hline
        Model & $n_{b}~(10^{-4}~{\rm fm^{-3}})$ & $P~(10^{-4}~{\rm MeV~fm^{-3}})$ & $\mathcal{E}~(10^{-1}~{\rm MeV~fm^{-3}})$ & Nucleus & $\mu_{e}~(\rm MeV)$ & $B/A$ (MeV)\\
        \hline
        DD--ME2 &  2.570 & 4.777 & 2.410 & $^{126}$Sr & 26.004 & 7.050\\
        DD--PC1 &  2.648 & 5.060 & 2.483 & $^{118}$Kr & 26.371 & 7.252\\
        DD--PCX &  2.444 & 4.768 & 2.292 & $^{120}$Sr & 25.992 & 7.398\\
        ELMA &  2.494 & 4.792 & 2.339 & $^{122}$Sr & 26.024 & 7.287\\
        \hline
    \end{tabular}
    \tablefoot{We list the baryon number density $n_b$, pressure $P$, energy density $\mathcal{E}$, last bound nucleus, electron chemical potential $\mu_e$, and binding energy per nucleon $B/A$.}
        \label{tab:table1}
\end{table*}

Figure~\ref{fig:NZ_vs_nb} shows the equilibrium proton and neutron numbers, $Z$ and $N$, as functions of the baryon density for the four nuclear mass models considered. Whenever available, experimentally known masses were adopted from AME2020~\citep{Huang_2021,Wang_2021}. The predicted compositions are nearly identical, while the equilibrium sequences remain within the experimentally constrained region. In this density range, the equilibrium sequence begins with $^{56}\mathrm{Fe}$ and proceeds through progressively more neutron-rich nuclei, including $^{62\text{--}64}\mathrm{Ni}$, $^{86}\mathrm{Kr}$, $^{84}\mathrm{Se}$, and $^{82}\mathrm{Ge}$.

Model-dependent differences first appear near the boundary of the experimentally constrained region. DD--PCX and ELMA predict $^{80}\mathrm{Zn}$ as the last equilibrium nuclide with an experimentally known mass, whereas the DD--ME2 sequence additionally includes $^{82}\mathrm{Zn}$. This isotope is of particular interest because its experimental mass measurement was shown to disfavour its presence in the outer crust, leading instead to a transition from $^{80}\mathrm{Zn}$ to the doubly magic nucleus $^{78}\mathrm{Ni}$~\citep{PhysRevLett.110.041101}. This behaviour is reproduced by DD--PCX and ELMA, while DD--ME2 remains the only case in which $^{82}\mathrm{Zn}$ is favoured over $^{78}\mathrm{Ni}$.

\begin{figure}
    \centering
    \includegraphics[width=\columnwidth]{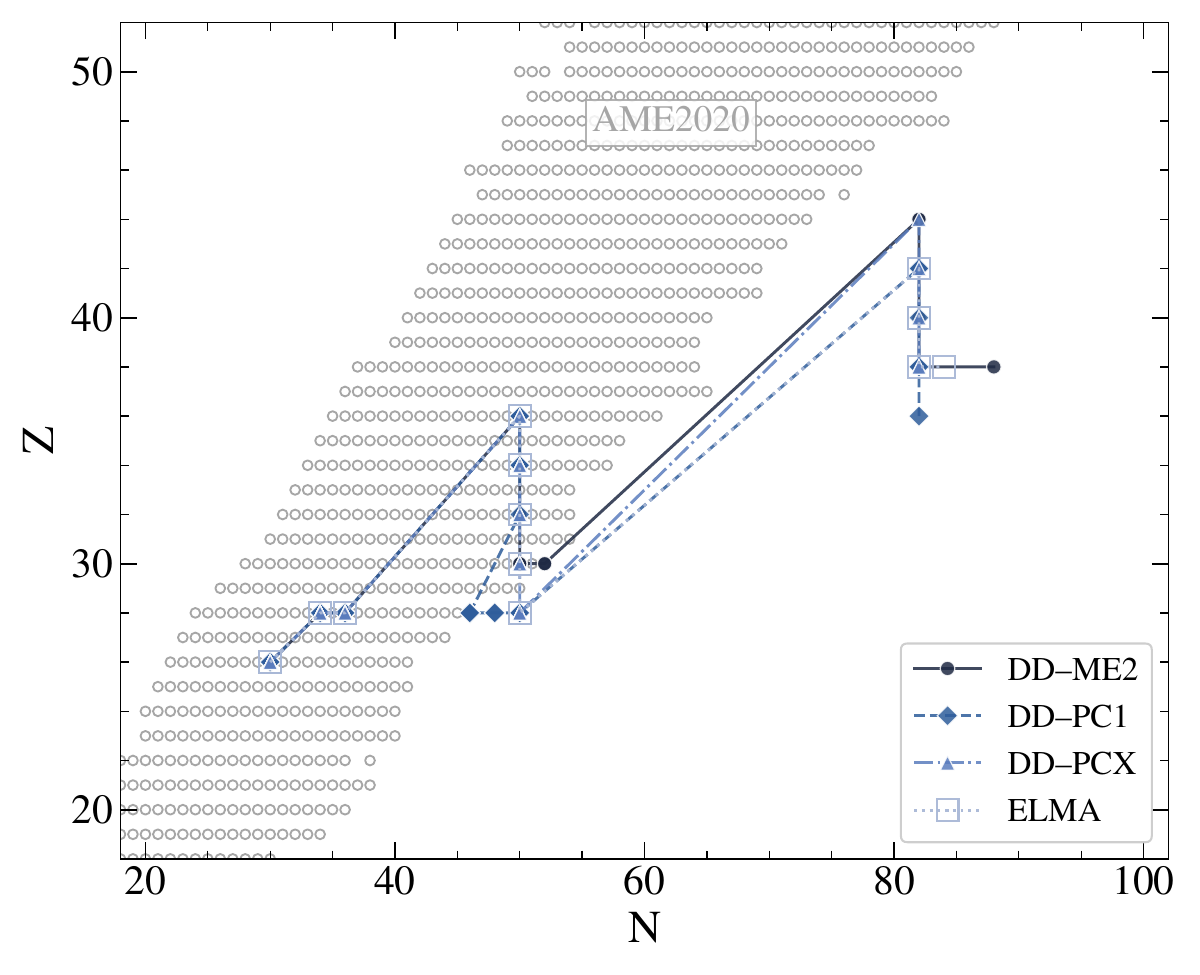}
    \caption{Equilibrium sequences of nuclei predicted by the four nuclear mass models in the $Z$--$N$ plane. The open circles indicate nuclei with experimentally known masses from AME2020~\citep{Huang_2021,Wang_2021}. The filled symbols connected by lines show the sequences obtained with the relativistic EDF mass tables, including their theoretical extrapolations towards neutron-rich nuclei, while the open squares indicate the ELMA sequence.}
    \label{fig:N_vs_Z}
\end{figure}

Deeper in the outer crust, the equilibrium sequences extend beyond the region of experimentally known masses and are therefore governed by theoretical mass extrapolations. The selection of $^{78}\mathrm{Ni}$ by three of the four models reflects the stabilising role of the $Z=28$ and $N=50$ shell closures. At still higher densities, all models predict an extended sequence of nuclei with the magic neutron number $N=82$, showing that shell effects remain important up to the vicinity of neutron drip~\citep{Baym-71,Haensel_1989,PhysRevC.83.065810}. This shell-driven structure is also visible in the complementary $Z$--$N$ representation shown in Fig.~\ref{fig:N_vs_Z}, where the equilibrium sequences cluster along the $N=50$ and $N=82$ shell closures before reaching the neutron-drip region. The extent of the $N=82$ sequence, however, differs among the models. DD--PC1 and DD--PCX remain at this shell closure up to the neutron drip, whereas DD--ME2 and ELMA extend beyond it, reaching maximum neutron numbers of $N=88$ and $N=84$, respectively.

\begin{figure}
    \centering
    \includegraphics[width=\columnwidth]{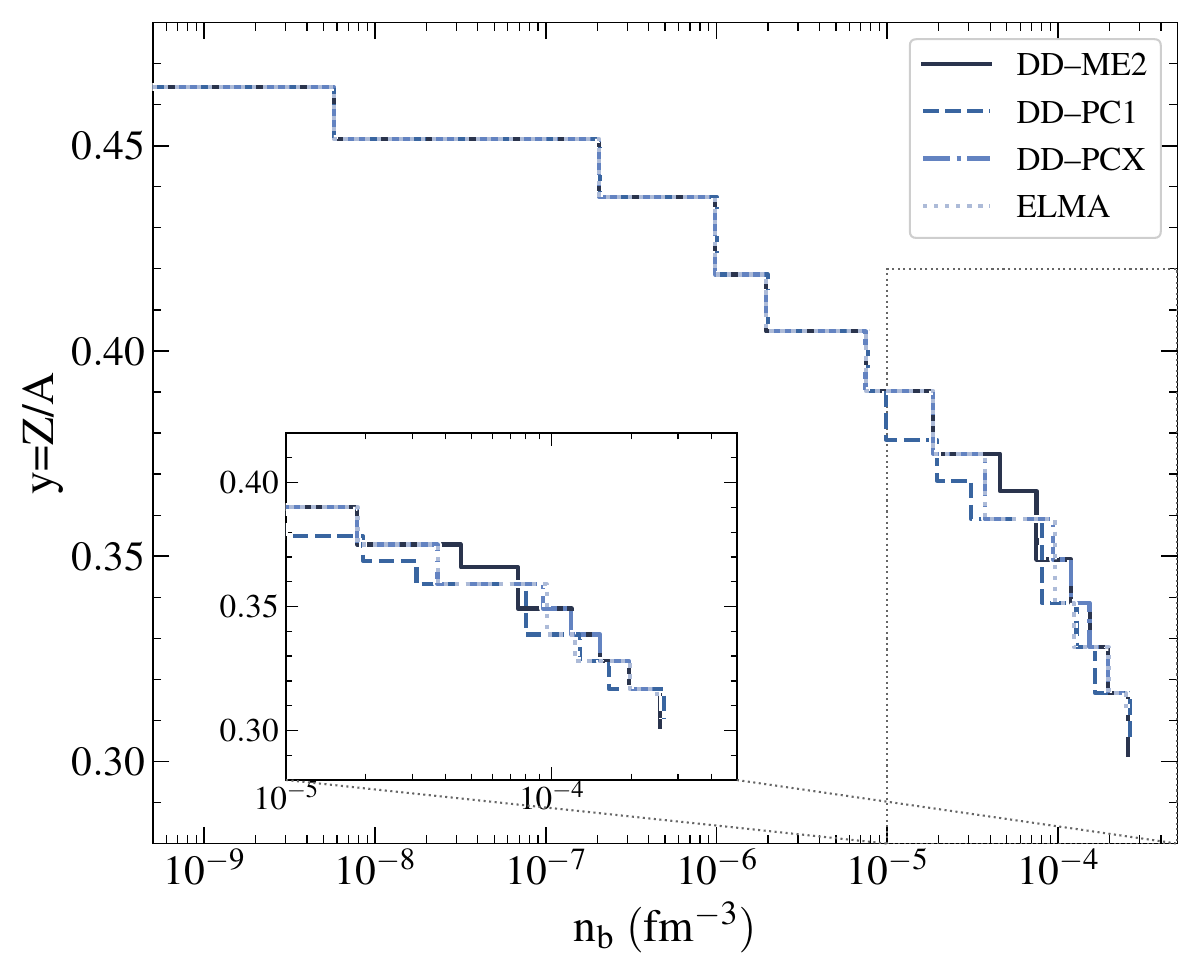}
    \caption{Proton fraction $y=Z/A$ as function of baryon density $n_b$ for the four nuclear mass models considered. The inset emphasises the region near the neutron-drip density.}
    \label{fig:yp_vs_nb}
\end{figure}

The DD--PC1 sequence exhibits a distinct behaviour just beyond the experimentally constrained region. In this model, the neutron number of the equilibrium nucleus decreases over a limited density interval before increasing at higher densities. However, this local behaviour does not indicate a reversal of the neutronisation of matter, since the proton fraction, $y=Z/A$, shown in Fig.~\ref{fig:yp_vs_nb}, decreases with increasing baryon density for all models, including DD--PC1. This feature reflects the pronounced sensitivity of the equilibrium nuclide sequence to small variations in the predicted nuclear binding energies, such that at zero temperature, even infinitesimal differences can alter the predicted composition~\citep{PhysRevC.83.065810}. It should be noted that the DD--PC1 interaction was calibrated primarily to the binding energies of medium-heavy and heavy nuclei in the mass regions $A\approx150\text{--}180$ and $A\approx230\text{--}250$~\citep{PhysRevC.78.034318}. Therefore, it is not specifically optimised for lower-mass regions, particularly those with $A\lesssim 90$ relevant to the outer-crust nuclei, whereas the other interactions considered here include calibration data spanning a much broader range from light to heavy nuclei.

The final segments of the equilibrium sequences, shown in the insets of Fig.~\ref{fig:NZ_vs_nb}, are dominated by Mo, Zr, and Sr isotopes, with Ru also appearing in DD--ME2 and DD--PCX and Kr appearing only in DD--PC1. With the exception of DD--PC1, the last equilibrium nuclide before the neutron drip is a neutron-rich Sr isotope. DD--PC1 instead terminates with $^{118}\mathrm{Kr}$, shifting the final composition to a lower-$Z$ isotopic chain.

The neutron-drip properties obtained with the four nuclear mass models are summarised in Table~\ref{tab:table1}. The largest model-to-model variations occur in the neutron-drip baryon density and energy density, both with relative spreads of approximately $8\%$. DD--PC1 predicts the highest neutron-drip density, $n_b=2.648\times10^{-4}~\mathrm{fm}^{-3}$, whereas DD--PCX gives the lowest value, $n_b=2.444\times10^{-4}~\mathrm{fm}^{-3}$. The pressure varies by about $6\%$, while the binding energy per nucleon of the terminal nuclei varies by approximately $5\%$. The electron chemical potential is the least model-dependent quantity, changing by only about $1.5\%$ across the models. Therefore, the mass model mainly affects the density and composition at which the neutron drip occurs, while the associated pressure and electron chemical potential remain comparatively stable because they are dominated by the degenerate electron gas. It is worth noting that the composition layers and corresponding thermodynamic quantities of the outer-crust EOSs remain close to that obtained in calculation of different outer-crust EOSs~\citep{Baym-71,Haensel_1989,Haensel_1994,PhysRevC.83.065810,10.1093/mnras/sty2413,Sharma_2015}. 

\begin{figure}
    \centering
    \includegraphics[width=\columnwidth]{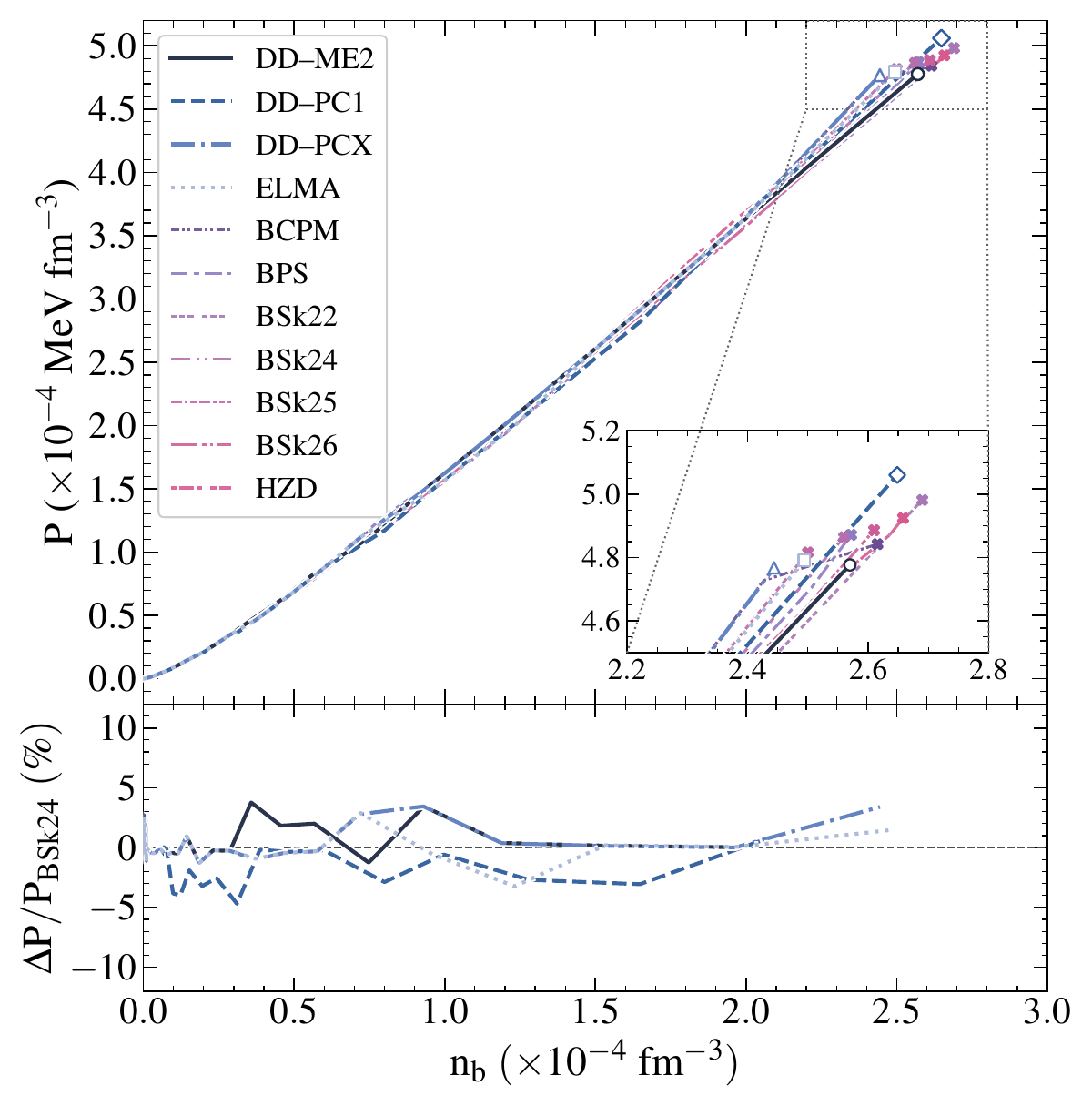}
    \caption{Pressure $P$ as a function of baryon density $n_b$ for the four nuclear mass models considered. The inset highlights the region near the neutron drip, with markers indicating the neutron-drip points. The BCPM~\citep{Sharma_2015}, BPS~\citep{Baym-71}, BSk22--BSk26~\citep{10.1093/mnras/sty2413}, and HZD~\citep{Haensel_1989} EOSs are shown for comparison. The lower panel displays the relative deviation with respect to the BSk24 EOS~\citep{10.1093/mnras/sty2413}.}
    \label{fig:p_vs_nb}
\end{figure}

Figure~\ref{fig:p_vs_nb} shows the pressure, $P$, as a function of baryon density for the four models considered. In agreement with the neutron-drip properties discussed above, the pressure curves remain close to one another throughout the outer crust. This reflects the dominant contribution of the degenerate electron gas, which makes the bulk EOS only weakly sensitive to the detailed equilibrium sequence. Model-dependent differences become visible mainly near the high-density end of the outer crust, where the composition is governed by extrapolated nuclear masses. In this region, DD--PCX gives the highest pressure at a fixed baryon density and therefore corresponds to the stiffest outer-crust EOS among the models, whereas DD--ME2 gives the lowest pressure and is the softest in the same interval.

For comparison, Fig.~\ref{fig:p_vs_nb} also shows the BCPM~\citep{Sharma_2015}, BPS~\citep{Baym-71}, BSk22--BSk26~\citep{10.1093/mnras/sty2413}, and HZD~\citep{Haensel_1989} EOSs. The present EOSs follow the same overall trend as these reference calculations throughout the outer-crust density range. The lower panel quantifies this comparison by showing the relative difference with respect to BSk24 EOS. BSk24 was selected as a representative EOS of the modern BSk family, since it belongs to the BSk22--BSk25 series fitted to the AME2012 nuclear mass evaluation, constrained to a relatively stiff neutron-matter EOS, and characterised by a symmetry energy of $J=30$ MeV, a value within the current range constrained by nuclear ground-state and excitation properties~\citep{ROCAMAZA201896,Koliogiannis_2026}.
The deviations remain at the level of $\sim 5\%$, with similar results obtained when the other reference EOSs are used as baselines. This confirms that the differences in the predicted equilibrium composition have only a limited impact on the bulk outer-crust EOS. The resulting outer-crust EOS tables are publicly available on Zenodo~\citep{outer_crust_eos}.

\begin{figure}
    \centering
    \includegraphics[width=\columnwidth]{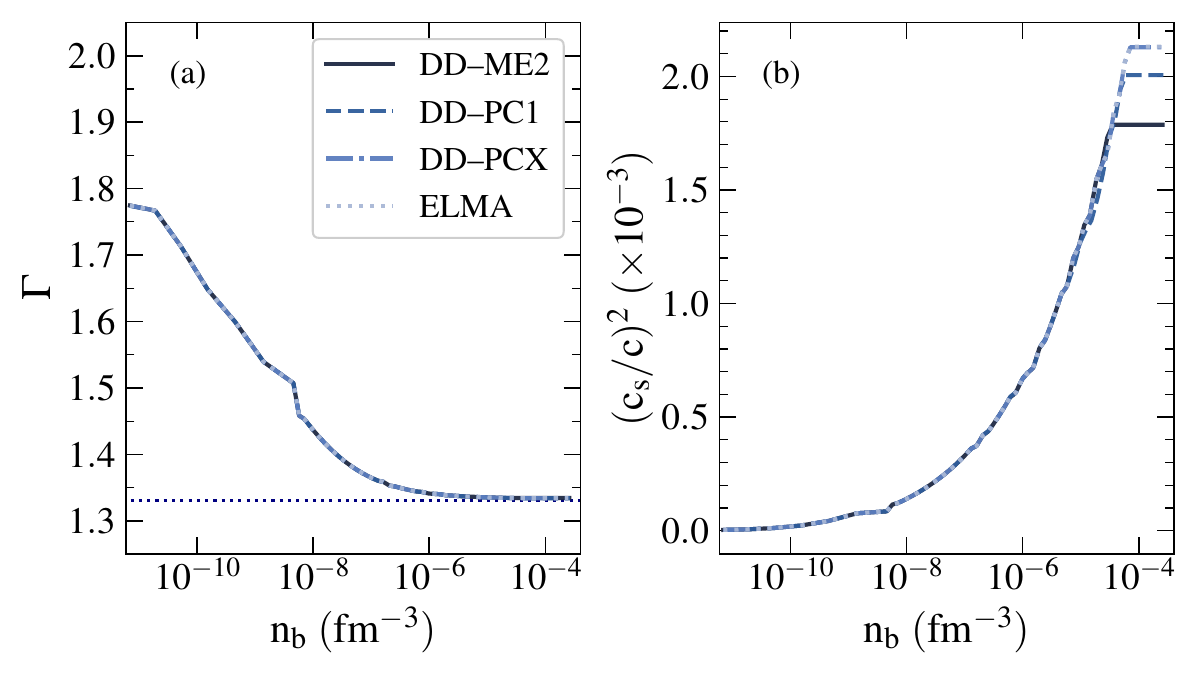}
    \caption{(a) Adiabatic index $\Gamma$ and (b) the square speed of sound in units of the speed of light $(c_s/c)^{2}$ as functions of the baryon density $n_b$ for the four nuclear mass models considered. The dotted horizontal line indicates the limit $\Gamma = 4/3$.}
    \label{fig:ad_cs}
\end{figure}

In addition to the pressure, the adiabatic index and the speed of sound provide complementary information on the local stiffness of the EOS. These quantities are defined as
\begin{equation}
    \Gamma = \frac{\partial \ln P}{\partial \ln n_b}\Bigg\vert_S, \quad \text{and} \quad \frac{c_s}{c} = \sqrt{\frac{\partial P}{\partial \mathcal{E}}\Bigg\vert_S},
\end{equation}
where $P$ is the pressure, $n_b$ is the baryon number density, and $\mathcal{E}$ is the energy density. As shown in Fig.~\ref{fig:ad_cs}, $\Gamma$ and $c_s/c$ for the four EOSs both exhibit very similar behaviour. 

The adiabatic index decreases within each interval of fixed composition and approaches $\Gamma \simeq 4/3$ at high densities. This behaviour is expected because towards the neutron drip, the matter pressure is dominated by the ultra-relativistic degenerate electron gas, while the leading correction to the energy density comes from the Coulomb lattice term. These two contributions scale approximately as $n_{b}^{4/3}$~\citep{Chamel_2008}. As the density increases, this scaling becomes increasingly dominant and drives $\Gamma$ towards $4/3$. The small oscillations superimposed on this trend arise from discrete changes in the equilibrium composition, while all EOSs remain mechanically stable throughout the outer crust. The speed of sound increases monotonically with density and remains below the causality limit, $c_s<c$, throughout the outer crust. Although model differences are visible in the pressure near the neutron drip, the corresponding differences in $\Gamma$ and $c_s$ remain small, indicating that the macroscopic stiffness of the outer-crust EOS is only weakly affected by the details of the nuclear mass extrapolation.

\subsection{Minimum-mass configuration}
\label{sec:M_min}
One way to determine the stability of neutron star configurations is the turning-point criterion~\citep{Shapiro-1983},
\begin{equation}
    \frac{dM}{d\mathcal{E}_c}=0,
\end{equation}
where $M$ is the gravitational mass, and $\mathcal{E}_{c}$ is the central energy density, which defines the boundaries of stable equilibrium along a sequence of configurations. This criterion gives rise to both a maximum and a minimum stable gravitational mass; configurations lying beyond these extrema are dynamically unstable. We only focused on the minimum-mass configuration.

Although neutron stars close to the theoretical minimum mass are not expected to be commonly produced in nature, their properties are important for understanding the stability sequence of compact stars. Observationally, the lowest accurately measured neutron-star candidate is the $1.174\pm0.004~M_\odot$ companion in the eccentric compact binary PSR~J0453+1559~\citep{Martinez_2015}. Recent three-dimensional neutrino-driven supernova simulations obtained a minimum birth mass of $1.192~M_\odot$, showing that such low-mass neutron stars can be produced within standard core-collapse scenarios~\citep{PhysRevLett.134.071403}. These values indicate that neutron stars formed through standard evolutionary channels are expected to have masses of order $\gtrsim 1.1$--$1.2~M_\odot$, which is well above the theoretical minimum mass of cold catalysed configurations.

The significance of the minimum mass is twofold. First, for isolated neutron stars, it marks the limiting configuration for which cold catalysed matter can remain gravitationally bound and stable. Second, in binary neutron-star systems, particularly during mass transfer, a low-mass component can be driven towards this stability threshold, beyond which the star becomes unstable and can undergo rapid structural rearrangement~\citep{Colpi_1989,Haensel_2002,PhysRevC.104.025805}.

\begin{figure}
    \centering
    \includegraphics[width=\columnwidth]{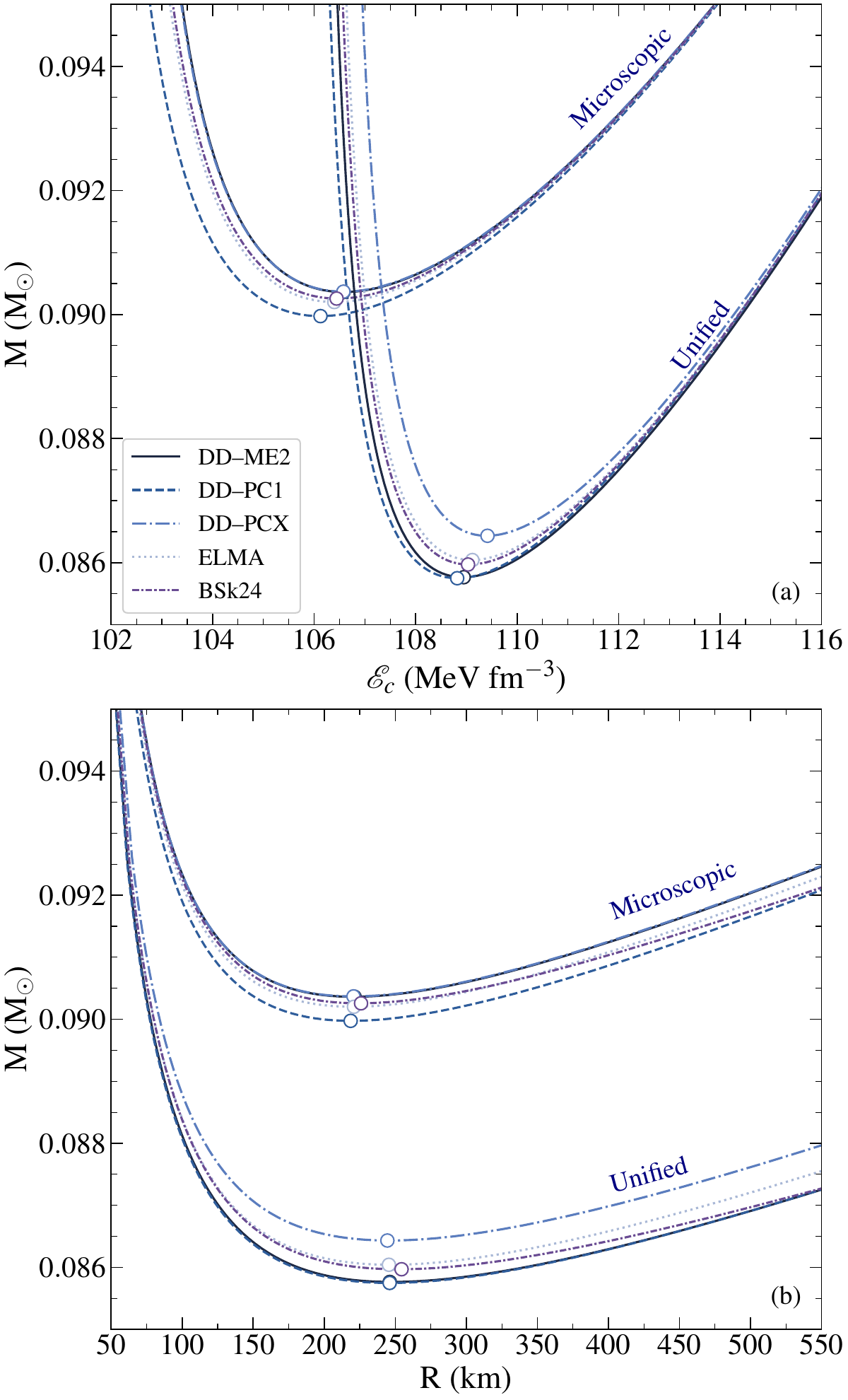}
    \caption{Gravitational mass $M$ as a function of (a) central energy density $\mathcal{E}_c$ and (b) radius $R$ for the considered outer- and inner-crust combinations, along with the DD--PC-J31 EOS for the liquid core. The circles denote the minimum-mass configuration. For comparison, the BSk24 EOS~\citep{10.1093/mnras/sty2413} is also shown.}
    \label{fig:mr}
\end{figure}

At the minimum-mass configuration, the most relevant bulk quantities are the minimum gravitational mass and the corresponding stellar radius. Previous studies have shown that the minimum mass is only weakly sensitive to the choice of the outer-crust EOS, with typical values clustered around $0.09~M_\odot$ when microscopic inner-crust models are employed~\citep{Baym-71,Haensel_2002}. Small model-to-model variations can nevertheless arise, since differences in the low-density EOS and in the crust--core matching prescription might propagate to the stellar structure. Figure~\ref{fig:mr} shows the gravitational mass $M$ as a function of (a) the central energy density $\mathcal{E}_c$ and (b) stellar radius $R$, for the four outer-crust EOSs considered. The markers indicate the minimum-mass configurations: (a) in the $M$--$\mathcal{E}_c$ plane, configurations to the left of the minimum are unstable, whereas those to the right are stable against infinitesimal radial perturbations; (b) in the $M$--$R$ plane, configurations to the right of the minimum are unstable, while those to the left are stable. Near the minimum mass, the curves present an extremely steep behaviour in the $M$--$\mathcal{E}_c$ plane, but form an extended nearly flat branch in the $M$--$R$ plane. This shows that small changes in the central energy density correspond to large variations in radius. This behaviour is a direct reflection of the internal structure of neutron stars, especially at sub-nuclear densities and near the minimum-mass configuration~\citep{Haensel_2002}.

\begin{table*}
    \centering
        \caption{\label{tab:table2}Properties of the minimum-mass neutron-star configurations for the considered EOS combinations.}
    \begin{tabular}{lllccccc}
        \hline\hline
        Outer crust & Inner crust & Liquid core & $n_{c}$ (fm$^{-3}$)  & $M_{\rm min}$ ($M_\odot$) & $R_{\rm min}$ (km) & $M_{\rm core}$ ($M_\odot$) & $R_{\rm core}$ (km)\\
        \hline
        DD--ME2 & Microscopic & DD--PC-J31 & 0.112 & 0.0904 & 221.29 & 0.0289 & 4.45 \\
        DD--PC1 &  &  & 0.111 & 0.0900 & 218.58 & 0.0283 & 4.42 \\
        DD--PCX &  &  & 0.112 & 0.0904 & 220.56 & 0.0289 & 4.45 \\
        ELMA &  &  & 0.112 & 0.0902 & 220.89 & 0.0287 & 4.44 \\
        BSk24 &  & & 0.112 & 0.0903 & 225.88 & 0.0287 & 4.44 \\
        \multicolumn{8}{c}{\vspace{-0.25cm}} \\
        DD--ME2 & Unified & DD--PC-J31 & 0.114 & 0.0858 & 246.12 & 0.0324 & 4.60 \\
        DD--PC1 &  &  & 0.114 & 0.0857 & 246.01 & 0.0322 & 4.59 \\
        DD--PCX &  &  & 0.115 & 0.0864 & 244.42 & 0.0331 & 4.63 \\
        ELMA &  &  & 0.114 & 0.0860 & 245.49 & 0.0327 & 4.61 \\
        BSk24 &  &  & 0.114 & 0.0860 & 254.40 & 0.0325 & 4.61 \\
        \hline
    \end{tabular}
    \tablefoot{Each combination comprises the outer-crust EOS, either the microscopic~\citep{Douchin_2001} or unified~\citep{PhysRevLett.83.3362,Carriere_2003} inner-crust EOS, and the DD--PC-J31 EOS~\citep{KOLIOGIANNIS2025139362,Koliogiannis_2026} for the liquid core. We list the central baryon density $n_c$, the minimum gravitational mass $M_{\rm min}$, the corresponding stellar radius $R_{\rm min}$, and the mass and radius of the liquid core, $M_{\rm core}$ and $R_{\rm core}$, respectively. The unified BSk24 EOS~\citep{10.1093/mnras/sty2413} is included for comparison.}
\end{table*}

The properties of the minimum-mass configurations are summarised in Table~\ref{tab:table2}. For the microscopic inner-crust EOS, the four outer-crust models give a minimum mass of $M_{\rm min}=0.0902 \pm 0.0002~M_{\odot}$ and a radius of $R_{\rm min}=220.33 \pm 1.35~\mathrm{km}$, with relative spreads 0.44\% and 1.23\%, respectively. When the unified inner-crust treatment is used, the minimum-mass configuration shifts to a slightly lower mass and a larger radius, with values $M_{\rm min}=0.0860 \pm 0.0003~M_{\odot}$ and $R_{\rm min}=245.51 \pm 0.85~\mathrm{km}$, with relative spreads 0.80\% and 0.70\%, respectively. The small scatter around these averages shows that the minimum-mass configuration is only weakly affected by the choice of outer-crust mass model. By contrast, the treatment of the inner crust has a more pronounced effect, lowering the minimum mass by $\sim5\%$ and increasing the radius by $\sim11\%$, corresponding to an increase of approximately $25~\mathrm{km}$.

The comparison therefore indicates that the dominant uncertainty in the minimum-mass configuration arises from the inner-crust EOS. However, although the effect of the outer-crust EOS is comparably weaker, it is not negligible: it affects the detailed layer structure and contributes to the precise values of the minimum mass and radius. An accurate treatment of the outer crust is therefore still required for a consistent description of low-mass neutron-star configurations.

\subsection{Crustal thickness and crustal moment of inertia}
\label{sec:crustal_moi}
\begin{figure}
    \centering
    \includegraphics[width=\columnwidth]{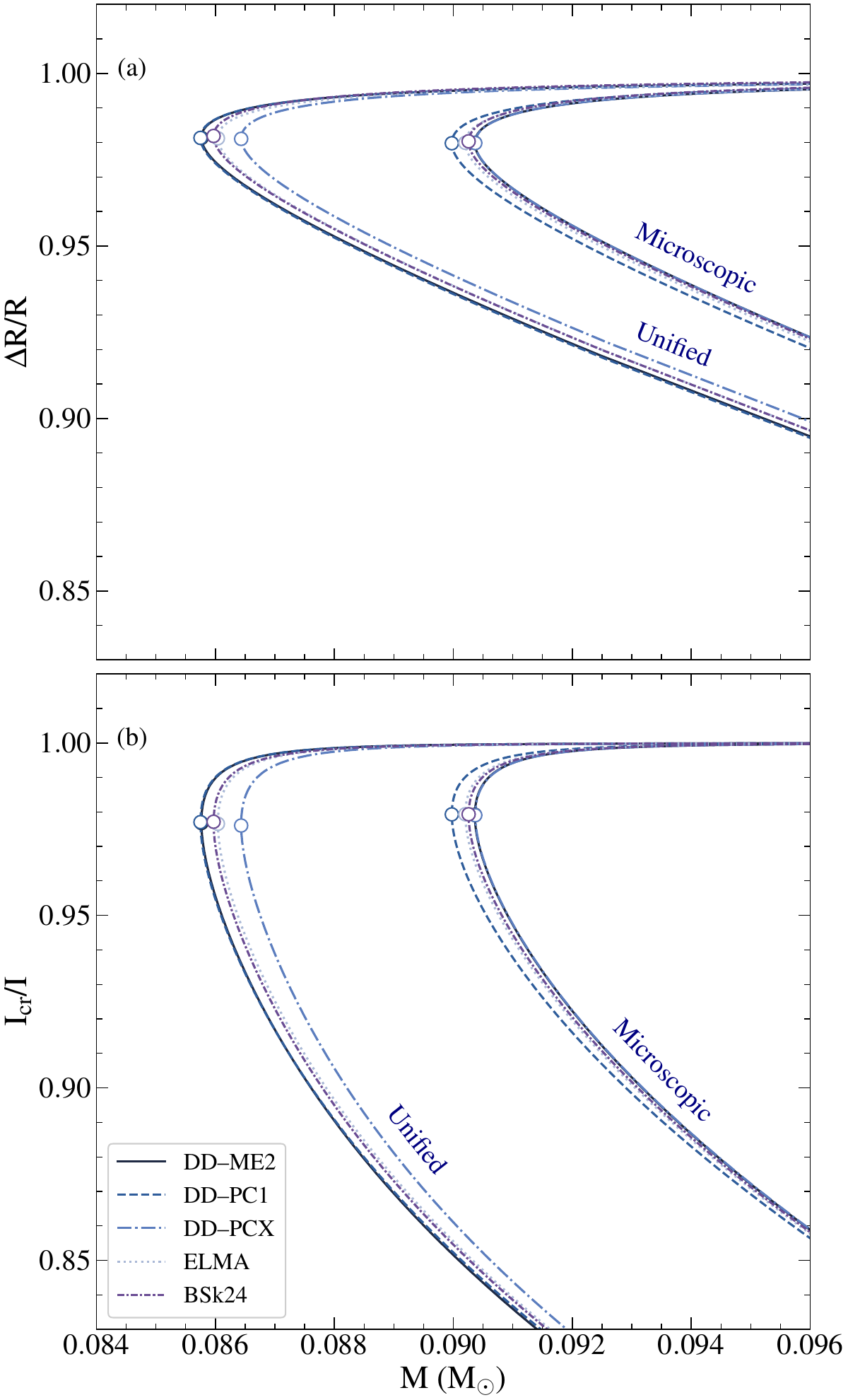}
    \caption{(a) Fraction $\Delta R/R$ and (b) fraction $I_{\rm cr}/I$ as functions of gravitational mass $M$ for the considered outer- and inner-crust combinations, along with the DD--PC-J31 EOS for the liquid core. The circles indicate the minimum-mass configuration. For comparison, the BSk24 EOS~\citep{10.1093/mnras/sty2413} is also shown.}
    \label{fig:cr_moi}
\end{figure}

The structural and rotational properties of neutron stars close to the minimum-mass configuration are mainly controlled by the extended crust~\citep{Baym-71,Xu_2009,Sharma_2015,PhysRevC.104.015801,PhysRevC.104.025805,Jiang_2025}. The radial extent of the crust is quantified by the crustal thickness, $\Delta R = R - R_{\rm core}$, where $R$ is the total stellar radius, and $R_{\rm core}$ denotes the radius of the liquid core. For the configurations considered here, the liquid core accounts for only $\sim 2\%$ of the total radius, emphasising the crust-dominated nature of minimum-mass neutron stars. The corresponding results are summarised in Tables~\ref{tab:table2} and~\ref{tab:table3}. The ratio $\Delta R/R$ is shown as a function of gravitational mass in Fig.~\ref{fig:cr_moi}a. It remains very close to unity, with $\Delta R/R\simeq 0.98$, and exhibits only small variations among the different outer-crust models. This shows that the global crustal geometry of minimum-mass configurations is only weakly affected by differences in the outer-crust composition.

\begin{table}
    \centering
        \caption{\label{tab:table3}Crustal properties of the minimum-mass neutron-star configurations for the considered EOS combinations.}
    \begin{tabular}{llccc}
        \hline\hline
        Outer crust & Inner crust & $\Delta R$ (km) & $\Delta R/R$ & $I_{\rm cr}/I$ \\
        \hline
        DD--ME2 & Microscopic & 216.84 & 0.979 & 0.979 \\
        DD--PC1 &  & 214.16 & 0.979 & 0.979 \\
        DD--PCX &  & 216.11 & 0.979 & 0.979 \\
        ELMA &  & 216.45 & 0.979 & 0.979 \\
        BSk24 & & 221.44 & 0.980 & 0.979 \\
        \multicolumn{5}{c}{\vspace{-0.25cm}} \\
        DD--ME2 & Unified & 241.52 & 0.981 & 0.976 \\
        DD--PC1 &  & 241.42 & 0.981 & 0.977 \\
        DD--PCX &  & 239.78 & 0.981 & 0.976 \\
        ELMA &  & 240.87 & 0.981 & 0.976 \\
        BSk24 & & 249.79 & 0.981 & 0.977 \\
        \hline
    \end{tabular}
    \tablefoot{The configurations employ the same outer-crust, inner-crust, and liquid-core EOS combinations as in Table~\ref{tab:table2}. We list the crustal thickness $\Delta R$, its fractional value $\Delta R/R$, and the fractional crustal moment of inertia $I_{\rm cr}/I$. The unified BSk24 EOS~\citep{10.1093/mnras/sty2413} is included for comparison.}
\end{table}

A complementary measure of the crustal contribution is the fractional crustal moment of inertia, $I_{\rm cr}/I$, where $I_{\rm cr}$ and $I$ denote the crustal and total moments of inertia, respectively. This quantity is relevant for the rotational evolution of neutron stars, in particular in connection with pulsar glitches, which are commonly interpreted as sudden angular-momentum transfer from the inner-crust neutron superfluid to the rigid crustal component~\citep{ANDERSON1975,PhysRevLett.83.3362,PhysRevLett.109.241103,PhysRevLett.110.011101}. As shown in Fig.~\ref{fig:cr_moi}b, $I_{\rm cr}/I$ is also close to unity, with $I_{\rm cr}/I\simeq 0.98$, and follows the same weak model dependence as the crustal-thickness ratio. The choice of inner-crust treatment also produces only very small changes in the fractional quantities $\Delta R/R$ and $I_{\rm cr}/I$. This behaviour is consistent with previous studies, which showed that crustal uncertainties have a limited effect on macroscopic neutron-star properties~\citep{PhysRevC.104.015801,PhysRevC.104.025805}, and that the fractional crustal moment of inertia depends only weakly on the adopted outer-crust EOS~\citep{Jiang_2025}. The agreement with these calculations, together with the small spread among the present models, provides an additional consistency check for the outer-crust EOSs constructed in this work.

\section{Conclusions}
\label{sec:concluding_remarks}
We presented a systematic assessment of the impact of nuclear-mass predictions far from stability on the construction of neutron-star outer-crust EOSs. Four outer-crust EOSs were constructed using contemporary nuclear mass inputs: the relativistic EDF-based mass tables DD--ME2, DD--PC1, and DD--PCX, together with the machine-learning-based mass model ELMA. Experimental masses from AME2020 were adopted wherever available, thereby restricting the influence of theoretical mass models on the neutron-rich region approaching the neutron drip.

The adopted mass models lead to differences in the final segments of the equilibrium nuclide sequence, in the identity of the last bound nuclei, and in the predicted neutron-drip density. These variations provide a measure of the residual uncertainty associated with extrapolating nuclear masses towards increasingly neutron-rich systems. Nevertheless, the corresponding thermodynamic properties remain closely aligned across all models, indicating that the model dependence is primarily reflected in the detailed crustal composition and in the precise location of the neutron drip, rather than in substantial modifications to the outer-crust EOS.

When implemented in neutron-star configurations near the minimum-mass limit, the four outer-crust EOSs yielded closely consistent predictions for the relevant stellar observables. The minimum gravitational mass, corresponding radius, crustal thickness, and fractional crustal moment of inertia exhibit only a weak dependence on the adopted outer-crust mass model. The use of two complementary inner-crust prescriptions mainly shifted the absolute values of the minimum mass and radius while preserving the overall conclusion that these stellar observables are largely insensitive to the choice of outer-crust EOS. The close agreement between the results obtained using ELMA and the relativistic EDF-based mass tables in the predicted equilibrium composition and in the related stellar properties further supports the use of machine-learning-based mass predictions in astrophysical applications involving nuclei beyond current experimental reach. The resulting outer-crust EOSs therefore provide reliable input for stellar modelling and a low-density benchmark for future development of unified neutron-star EOSs.

\section*{Data availability}
The outer-crust EOS tables generated in this work are publicly available on Zenodo at~\citet{outer_crust_eos}.

\begin{acknowledgements}
This work is supported by the Croatian Science Foundation under the project number HRZZ-MOBDOL-12-2023-6026 and under the project Relativistic Nuclear Many-Body Theory in the Multimessenger Observation Era (HRZZ-IP-2022-10-7773).
This work was supported by the project ``Implementation of cutting-edge research and its application as part of the Scientific Center of Excellence for Quantum and Complex Systems, and Representations of Lie Algebras'', Grant No. PK.1.1.10.0004, co-financed by the European Union through the European Regional Development Fund - Competitiveness and Cohesion Programme 2021-2027. This paper was supported by the European Union – NextGenerationEU through the National Recovery and Resilience Plan 2021-2026 -- Institutional grant of University of Zagreb Faculty of Science (Nuclear Astrophysics). This research was supported by the European Union – NextGenerationEU through the National Recovery and Resilience Plan 2021–2026 Institutional grants of University of Zagreb Faculty of Science (PMF-PRESTIGE).
\end{acknowledgements}

\bibliographystyle{aa}
\bibliography{aa_bib}

\end{document}